\documentclass[12pt,a4paper]{article}

\usepackage{amsmath}
\usepackage{amsfonts}
\usepackage{amssymb}
\usepackage{graphicx,epsfig}

\begin{document}

\title{Theory of polyelectrolyte dendrigrafts}
\author{Oleg V.Borisov$^{1,2,3}$, Oleg V.Shavykin$^{3}$, 
Ekaterina B.Zhulina$^{1,2}$, \\
$^1$ Institut des Sciences Analytiques et de Physico-Chimie\\
pour l'Environnement et les Mat\'eriaux,\\ 
CNRS, Universit\'e de Pau et des Pays de l'Adour UMR 5254,\\
64053 Pau, France\\
$^2$Institute of Macromolecular Compounds \\
of the Russian Academy of Sciences, \\
199004 St.Petersburg, Russia \\
$^3$Saint Petersburg National Research University of Information \\
Technologies, Mechanics and Optics, 197101 St.Petersburg, Russia \\
}
\date{20 November, 2019}
\maketitle

\begin{abstract}
A mean field approach is used to analyze equilibrium conformations of
polyelectrolyte dendrigrafts comprising ionically charged dendrons attached
by focal points to flexible linear backbone. Power law dependences for local
structural parameters: cross-sectional thickness and intergraft distance,
are derived as a function of grafting density and degree of branching of the
dendrons. The cases of quenched and pH-sensitive ionization of the dendrons
are considered. The finite extensibility of the backbone is taken into
account. It is demonstrated that an increase in the degree of branching of the dendrons
leads to a decrease in the
dendrigraft thickness compared to that of the polyelectrolyte
molecular brush  with the same degree
of polymerization of the side chains, while intergraft distance either increases or stays close
to counter length of fully extended backbone spacer. 
The analytical mean-field theory predictions are confirmed by results of numerical self-consistent field modelling.
\end{abstract}
\vspace{0.5cm}
{\it Keywords:}\hspace{0.3cm} dendrimers, polyelectrolytes, mean-field theory

\newpage

\section{Introduction}

Dendrimers or hyperbranched polymers are currently actively explored as
nano-carriers for anticancer drugs or siRNA in anti-cancer therapies since
they offer sufficient loading capacity for cargo combined with multiple
terminal groups assessable for modifications with required smart
functionalities (e.g for targeted delivery)\cite{Dendrimers,Dendrimers_Medicine}. 
Cationic poly(amidoamine) (PAMAM) or poly-L-lysine
dendrimers are the best known examples.

The higher generation dendrons exhibits therefore the best performance,
though they are synthetically very demanding. At least two alternatives were
suggested: The first is to use so-called supramolecular dendrimers, or
dendromicelles, that are the outcome of spontaneous self-assembly of
amphiphilic macromolecules comprising associating hydrophobic linear block
covalently linked to a dendron of the 2-3 generation \cite%
{hybrids_7,hybrids_3}. Since a single dendromicelles may comprise tens or
even hundreds of dendrons exposed to the external solution, it can
overperform regular dendrimers of higher generation in drug or siRNA
delivery \cite{Peng_1,Peng_2}

Another approach is to use dendrigrafts that comprise multiple (tens to
hundreds) dendrons covalently attached through their focal points to a
nano-colloidal particle or to a linear backbone\cite{Teertstra}. Similar to
graft copolymer of molecular brushes dendrigrafts can be prepared by
grafting to, grafting from or grafting through methods, that are nowadays
well elaborated \cite{Teertstra,Axel}.

There are multiple reports on therapeutic effect of cationic dendrimers and
dendrigrafts in preventing formation of amyloid peptide assemblies
responsible for neuro-degenerative diseases \cite%
{amyloids_1,amyloids_2,amyloids_3}

Conformational properties of non-ionic dendronized polymers which
architecturally resemble dendrigrafts were studied recently by means of
scaling and self-consistent field theories \cite{dendronized_polymer_1,
dendronized_polymer_2,dendronized_polymer_3}. Ionically charged dendrigrafts
can be viewed as molecular polyelectrolyte brushes with dendritically
branched ionically charged side chains, hence, they exhibit many features
inherent for polyelectrolyte brushes \cite{Ballauff,Ruhe2004}, in
particular, capability to accumulate mobile counterions neutralizing ionic
charge of the dendrons in the intramolecular volume of the dendrigraft. To
account for the effect of the counterion localization and its consequences
for the polyelectrolyte brush or dendrigraft conformation, a non-linear
Poisson-Boltzmann approach has to be employed.

In the present study we focus on the effect of ionic interactions that
govern conformations of most practically relevant cationic dendrigrafts
which were not explored so far.

\section{Model of a polyelectrolyte dendrigraft and mean-field formalism}

\subsection{Dendrigraft model}

The dendrigraft macromolecule is formed by multiple ionically charged
dendrons attached at regular intervals by their focal points (terminal
segments of the root spacers) to the linear chain backbone,\textbf{Figure 1}. Each dendron is
characterized by the number of generations, $g=0,1,2,...$, functionality of
branching points $q=1,2,3,...$ and number of monomer units in one (flexible)
spacer $n$. Total number of monomer units per dendron equals 
\begin{equation}
N=n(q^{g+1}-1)/(q-1)  \label{N}
\end{equation}

\begin{figure}[ht]
\center{\includegraphics[scale=0.50]{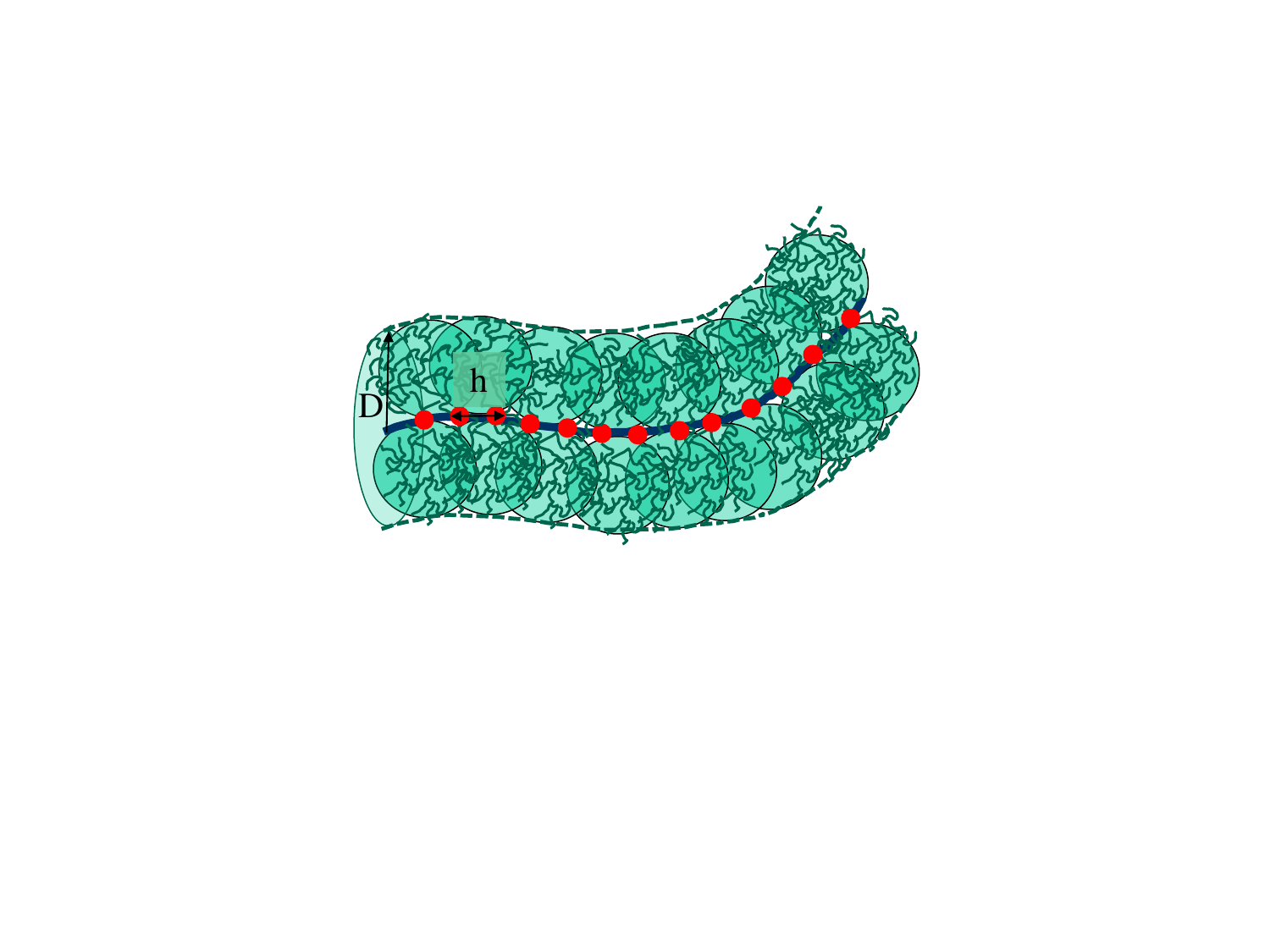}}
\caption{Schematic of a dendrigraft. Red circles indicate grafting (focal) points of dendrons , 
charges and counterions are not shown.
}
\end{figure}

The number of dendrons in the dendrigraft is $P\approx M/m\gg 1$ where $M$
is total number of monomer units in the backbone and $m$ is the number of
monomer units in a segment of the backbone separating two neightboring
grafting points. Typically $m$ is sufficiently small to
ensure crowding and strong interactions between neighboring dendrons. 
We assume both the backbone of the dendrigraft and spacers in the dendrons
to be flexible, with the monomer unit length $a$ on the order of the
statistical segment length. 

The fraction $\alpha$ of monomer units in the dendrons are elementary
(positively) charged so that the total charge of one dendron is $Q=\alpha N$%
. We assume that $\alpha\ll (a/l_{B})^2\cong 1$ where $l_{B}=e^{2}/k_{B}T%
\varepsilon $ is the Bjerrum length 
($e$ is the elementary charge, $\varepsilon $ is the dielectric constant of
the solvent, $T$ is the temperature and $k_{B}$ is the Boltzmann constant).
That is, weak electrostatic coupling regime takes place.

Below we distinguish cases of quenched polyelectrolytes, where $\alpha $ is
constant and independent of external conditions 
and annealing (or weak) polyelectrolytes where $\alpha $ depends on the
solution pH, ionic strength and, eventually, conformation of the dendrigraft.

We consider an isolated dendrigraft molecule in dilute aqueous solution
(buffer) which contains mobile (monovalent) counterions neutralizing the
charge of the dendrigraft and also monovalent salt. 
The salt concentration $\phi_s$ specifies the Debye screening length as $\kappa^{-1}=(4\pi l_{B}\phi_{s}/a)^{-1/2} $.
(Here and below we use dimensionless salt concentration, i.e. multiplied by $a^{3}$).

At low degree of ionization $\alpha$ or/and high salt concentration the
non-electrostatic excluded volume interactions between crowded dendrons may
come into play. We assume that water is marginally good or even close to
theta-solvent for uncharged monomer units of the dendrons which corresponds
to the dominance of three-body over two-body repulsions.

\subsection{Free energy}

Dendrigraft with sufficiently long backbone and large number of attached to
it dendrons can be assimilated to a cylindrical brush in which local
cylindrical symmetry and axial extension of the backbone are governed by
strong repulsive interactions between dendrons.

Local conformation of \ a dendrigraft is characterized by axial extension
(end-to-end distance) $h$ of the backbone segments (spacers) connecting
neighboring grafting points, and the cross-sectional thickness $D$ (i.e.,
typical distance between the focal points and terminal segments of the
dendrons). Below we use a mean-field approximation in which the gradient in
transverse distribution of monomer units in the dendrigraft is neglected,
and the volume fraction of monomer\ units in the cross-section is
characterized by its average value. That is, we introduce boxlike model for
cylindrical brush of dendrons encompassing the dendrigraft backbone. Within
this approximation the free energy (per dendron) can be presented as 
\begin{equation}
F(D,h)=F_{conf}^{(n)}(D)+F_{conf}^{(m)}(h)+F_{int}(D,h)  \label{F}
\end{equation}%
Here $F_{conf}^{(n)}$ and $F_{conf}^{(m)}(h)$ account for the
conformational entropy penalty for extension of the dendron in the direction
normal to the dendrigraft backbone and lateral elongation of the spacer,
respectively, while $F_{int}$ describes electrostatic and excluded volume
repulsions between monomer units.

Equilibrium local structural properties of the dendrigraft can be obtained
by minimization of the free energy, eq \ref{F}, with respect to $D$ and $h$,
or, equivalently, by equating the elastic tension arising due to repulsive
interactions between dendrons to the restoring entropic force acting in the
stretched backbone.

To evaluate $F_{conf}^{(n)}$ we assume the Gaussian elastic response of the
extended segments of dendrons and use the expression for boxlike model of
dendron brush \cite{Lebedeva2017}, 
\begin{equation}
F_{conf}^{(n)}(D)/k_{B}T\cong \frac{D^{2}}{Na^{2}}\eta ^{2}  \label{Fconf_D}
\end{equation}%
Here and below we omit the numerical pre-factors on the order of unity and
weak logarithmic dependences. The so-called topological ratio $\eta $ in eq %
\ref{Fconf_D} quantifies relative increase in the conformational free energy
penalty for stretching of (dendritically) branched macromolecule compared to
that for a linear chain with the same number $N$ of monomer units. Notably
eq \ref{Fconf_D} is consistent with a more accurate self-consistent field
description of dendron brushes with nonuniform transverse distribution of
monomer units around the backbone of dendrigraft (see discussion in ref \cite%
{Lebedeva2017}). The latter approach provides the values of $\eta $ for a
number of the graft architectures. \ In particular, for dendrons of the
first generation with $q$ free branches of $n$ monomer units each \cite%
{Macromolecules2012} 
\begin{equation}
\eta _{1}=\frac{2(q+1)}{\pi }\arctan \left( \frac{1}{\sqrt{q}}\right)
\approx \frac{2}{\pi }\sqrt{q}\text{\ \ if \ }q\gg 1  \label{n1}
\end{equation}%
while for dendrons of the second generation (with total number $%
N=n(1+q+q^{2})$ of monomer units) \cite{Macromolecules2015} 
\begin{equation}
\eta _{2}=\frac{2(q^{2}+q+1)}{\pi }\arctan \left( \frac{1}{\sqrt{q(2+q)}}%
\right) \approx \frac{2}{\pi }q\text{ \ if \ }q\gg 1\text{\ \ \ \ }
\label{n21}
\end{equation}

The values of $\eta $ for dendrons with larger number of generations can be
calculated analytically or numerically using the conditions of length
conservation for all chain segments between branching monomers, and balance
of the elastic forces in all branching points of the macromolecule \cite%
{Macromolecules2015}.

As prompted by eqs \ref{n1},\ref{n21}, in brushes of regular dendrons with
number of generation $g$, an asymptotic expression for $\eta $ is given by 
\begin{equation}
\eta \approx \frac{2}{\pi }q^{g/2}\text{\ \ \ \ \ if \ }q\gg 1\text{\ \ \ }
\label{eta}
\end{equation}

The repulsions between tethered dendrons may cause strong stretching of
spacers in the dendrigraft backbone. In order to account for non-linear
elasticity and limiting extensibility of the backbone, we use the equation
for elastic response of a freely-joint (ideal) chain on simple cubic lattice
(scl)\cite{Polotsky2010}:

\begin{equation}
\frac{f(h)}{k_{B}T}=-\frac{\partial F_{conf}^{(m)}(h)/k_{B}T}{\partial h}%
=a^{-1}\biggl(\ln (\frac{2x+\sqrt{1+3x^{2}}}{1-x})\biggr)_{x=h/ma}
\label{f_h}
\end{equation}%
to give 
\begin{equation}
\frac{f(h)}{k_{B}T}\approx \frac{3x}{a}=\frac{3h}{ma^{2}}\text{ \ if \ }%
x=h/ma\ll 1  \label{fhl}
\end{equation}%
in the linear elasticity regime, and logarithmic divergency at strong
extensions, 
\begin{equation*}
\frac{f(h)}{k_{B}T}\approx a^{-1}\ln \left( \frac{4}{1-x}\right) \text{ \ if
\ }x=h/ma\rightarrow 1
\end{equation*}%
thus assuring finite extensibility of spacers.

A particular form of the interaction free energy contribution, $F_{int}(D,h)$%
, depends on the extent of screening of intramolecular Coulomb interactions.
It should be specified separately for strong and weak polyelectrolytes.

\section{Strong polyelectrolyte dendrigraft (quenched charges)}

Depending on the extent of screening of intramolecular Coulomb interactions,
the following physical regimes with specific approximate expressions for $%
F_{int}(D,h)$ can be distinguished:

\subsection{Bare Coulomb repulsion (charged regime C)}

At low salt concentration $\phi_{s}$and small fraction $\alpha $ of
charged monomer units, bare Coulomb repulsions between dendrons with focal
points separated by distance $h$ along the backbone, lead to 
\begin{equation}
F_{int}(D,h)/k_{B}T\cong \frac{l_{B}(\alpha N)^{2}}{h}\ln \frac{Ph}{D}
\label{F_p}
\end{equation}
which applies at $Ph\gg D$.
Minimization of the free energy per dendron, $F_{conf}^{(n)}(D)+F_{int}(D,h)$%
,with respect to $D$ leads to 
\begin{equation}
D/a\cong N^{3/2}h^{-1/2}(\alpha ^{2}l_{B}/a)^{1/2}\eta ^{-1}  \label{D_p_1}
\end{equation}%
and with accuracy of logarithmic prefactor 
\begin{equation}
F_{conf}^{(n)}(D)+F_{int}(D,h)\cong \frac{l_{B}(\alpha N)^{2}}{h}  \label{Fd}
\end{equation}%
Subsequent minimization of $F$ in eq \ref{F} with respect to $h$ , 
\begin{equation*}
\frac{\partial \lbrack F_{conf}^{(n)}(D)+F_{int}(D,h)]}{\partial h}+f(h)=0
\end{equation*}%
with the account of eq \ref{Fd}, and linearized form (eq \ref{fhl}) of eq %
\ref{f_h}\ gives 
\begin{equation}
h/a\cong N^{2/3}(\alpha ^{2}l_{B}/a)^{1/3}m^{1/3}  \label{h_p}
\end{equation}%
which is remarkably independent of architecture of the dendrons (independent
of $\eta $).

Finally, using eq \ref{D_p_1} one finds thickness $D$ of the dendron
cross-section as 
\begin{equation}
D/a\cong N^{7/6}(\alpha ^{2}l_{B}/a)^{1/3}m^{-1/6}\eta ^{-1}  \label{D_p_2}
\end{equation}%
Using approximate expression for $\eta $ in eq \ref{eta} we find that at $%
n=const$ thickness $D$ of the dendrigraft increases with number $g$ of
generations as 
\begin{equation*}
D/a\sim q^{2g/3}
\end{equation*}

\subsection{Coulomb repulsions screened by counterions (osmotic regime O)}

Similar to cylindrical polyelectrolyte brushes, dendrigrafts with
sufficiently large number of charges per unit length of the backbone, $%
\alpha N/h$, accumulate in the intramolecular volume counterions that
neutralize bare charge of the dendrons even in salt-free solutions. The
onset of the counterion localization is estimated from the condition $\alpha
Nl_{B}/h\geq 1$ and, as follows from eq \ref{h_p} is independent of the
architectural parameters of the dendrons and controlled only by their bare
charge $\alpha N$. More specifically, counterions get localized within
dendrigrafts if $\alpha N/m\geq 1$ (and we assumed $l_{B}\cong a$).

At low salt concentration the interaction term in the free energy of the
dendrigraft is dominated by translational entropy of counterions localized
in the intra-molecular volume, 
\begin{equation}
F_{int}(D,h)/k_{B}T\cong \alpha N\ln (\alpha N/hD^{2})  \label{F_osm}
\end{equation}%
and minimization of the free energy with respect to $D$ leads to 
\begin{equation}
D/a\cong \alpha ^{1/2}N\eta ^{-1}  \label{D_osm}
\end{equation}%
which is remarkably independent of $m$ and strongly decreases as a function
of the degree of branching if $N$ is kept constant. 

The extensional force induced in the backbone due to interactions between
the dendrons in the osmotic regime 
\begin{equation}
\frac{\partial F_{int}(D,h)}{\partial h}\cong -\frac{\alpha N}{h}
\label{f_osm}
\end{equation}%
is also independent of the dendron's architecture. By balancing it with the
restoring force arising in the backbone given by eq \ref{f_h}, we obtain the
following expression for the size of the backbone spacer %

\begin{equation}  \label{h_osm}
\biggl\{x \ln(\frac{2x+\sqrt{1+3x^2}}{1-x}\biggr\}_{x=h/ma}=\frac{\alpha N}{m%
}
\end{equation}

Since the osmotic regime (localized counterions) takes place at $\alpha
N/m\gg 1$, eq \ref{h_osm} indicates that the spacers in the backbone are
stretched in the osmotic regime up to the limit of their extensibility, 
\begin{equation}
h/ma\approx 1-4\exp {(-\alpha N/m)}
\label{h_lim}
\end{equation}%
(More accurately, the crossover between the regimes of bare Coulomb
repulsion and osmotic regime, i.e. the counterion localization threshold,
occurs at $\alpha N/m\geq (a/l_{B})^{2})$, but we assume that $l_{B}\approx
a $).

If the length $n$ of
spacers in the dendrons is fixed, then, as follows from eqs \ref{D_osm} and \ref{eta}, the cross-sectional thickness of the
dendrigraft grows with as a function of number of generations in the
dendrons as
\begin{equation*}
D/a\sim q^{g/2}
\end{equation*}%
i.e. slightly stronger than in the regime of non-screened Coulomb repulsions.

\subsection{Coulomb repulsions screened by added salt (salt dominated regime
S)}

If low molecular weight salt is added to the solution at concentration
exceeding average concentration of counterions entrapped in the
intramolecular volume of the dendrigraft, then co- and counterions of salt
provide a dominant contribution to the screening of interactions between
charged dendrons. The threshold concentration of salt $\phi_{s}^{\ast }\cong
\eta ^{2}/(mN)$ is remarkably independent of $\alpha $ (which is a specific
feature of the cylindrically-symmetric distribution of charge density in
long dendrigraft), but strongly increases as a function of the degree of
branching, i.e., an increase in $\eta $.

Similarly to the salt-added semidilute polyelectrolyte solution, the
interaction free energy can be presented within the mean-field approximation
as

\begin{equation}
F_{int}(D,h)/k_{B}T\cong \alpha ^{2}l_{B}\kappa ^{-2}\frac{N^{2}}{hD^{2}}
\label{F_salt}
\end{equation}%
i.e., as an outcome of binary monomer-monomer interactions with the
effective salt-dependent second virial coefficient $\alpha ^{2}l_{B}\kappa
^{-2}$.

Minimization of the free energy with account of eq \ref{F_salt} leads to

\begin{equation}
D/a\cong N^{7/10}m^{-1/10}(\frac{\alpha ^{2}}{\phi_{s}})^{1/5}\eta ^{-3/5}
\label{D_s}
\end{equation}%
\begin{equation}
h/a\cong N^{1/5}m^{2/5}(\frac{\alpha ^{2}}{\phi_{s}})^{1/5}\eta ^{2/5}
\label{h_s}
\end{equation}%
and we used linearized form (eq \ref{fhl}) of eq \ref{f_h} because screening
of the inter-dendron repulsions upon addition of sufficient amount of salt
enables the spacers of the backbone to relax their extension with respect to
full stretching.

As we can see from eqs \ref{D_s}, \ref{h_s}, both $D$ and $h$ are decreasing
functions of salt concentration.

An increase in the degree of branching of the dendrons (at constant $N$)
leads to a decrease in $D$ but an increase in $h$.

If the number of generation in dendrons is increased upon keeping constant
the number of monomer units $n$ per spacer, then both the cross-section
thickness and the extension of spacers increase as 
\begin{equation*}
D/a\sim q^{2g/5}
\end{equation*}%
\begin{equation*}
h/a\sim q^{2g/5}
\end{equation*}

\subsection{Dendrigrafts dominated by non-electrostatic interactions
(quasi-neutral regime QN)}

At very high salt concentrations $\phi_{s}$ the ionic interactions are fully
screened off, and conformational properties of the dendrigraft are governed
by non-electrostatic (excluded volume) repulsions. Assuming that solvent is
close to theta-solvent for monomer units of the dendrigraft (dominance of
three-body over two-body repulsions and higher order interactions), one
presents the interaction free energy 
\begin{equation}
F_{int}(D,h)/k_{B}T\cong \frac{N^{3}}{(hD^{2})^{2}}  \label{F_quasineutr}
\end{equation}%
with the third virial coefficient of monomer-monomer interactions $%
w/a^{6}\cong 1$. Minimization of the free energy with respect to $D$ and $h$
leads to 
\begin{equation}
D/a\cong N^{5/8}m^{-1/8}\eta ^{-1/2}  \label{D_qn}
\end{equation}%
\begin{equation}
h/a\cong N^{1/8}m^{3/8}\eta ^{1/2}  \label{h_qn}
\end{equation}%
If $n=const$, then 
\begin{equation*}
D/a\sim q^{3g/8}
\end{equation*}
\begin{equation*}
h/a \sim q^{3g/8}
\end{equation*}

Notably, at small values of $m$ the backbone spacer remains strongly
stretched, $h/a\cong m$, and in this case dendrigraft thickness $D$ is
specified as 
\begin{equation}
D/a\cong N^{2/3}m^{-1/3}\eta ^{-1/3}  \label{D_qn_1}
\end{equation}

\begin{figure}[ht]
\center{\includegraphics[scale=0.50]{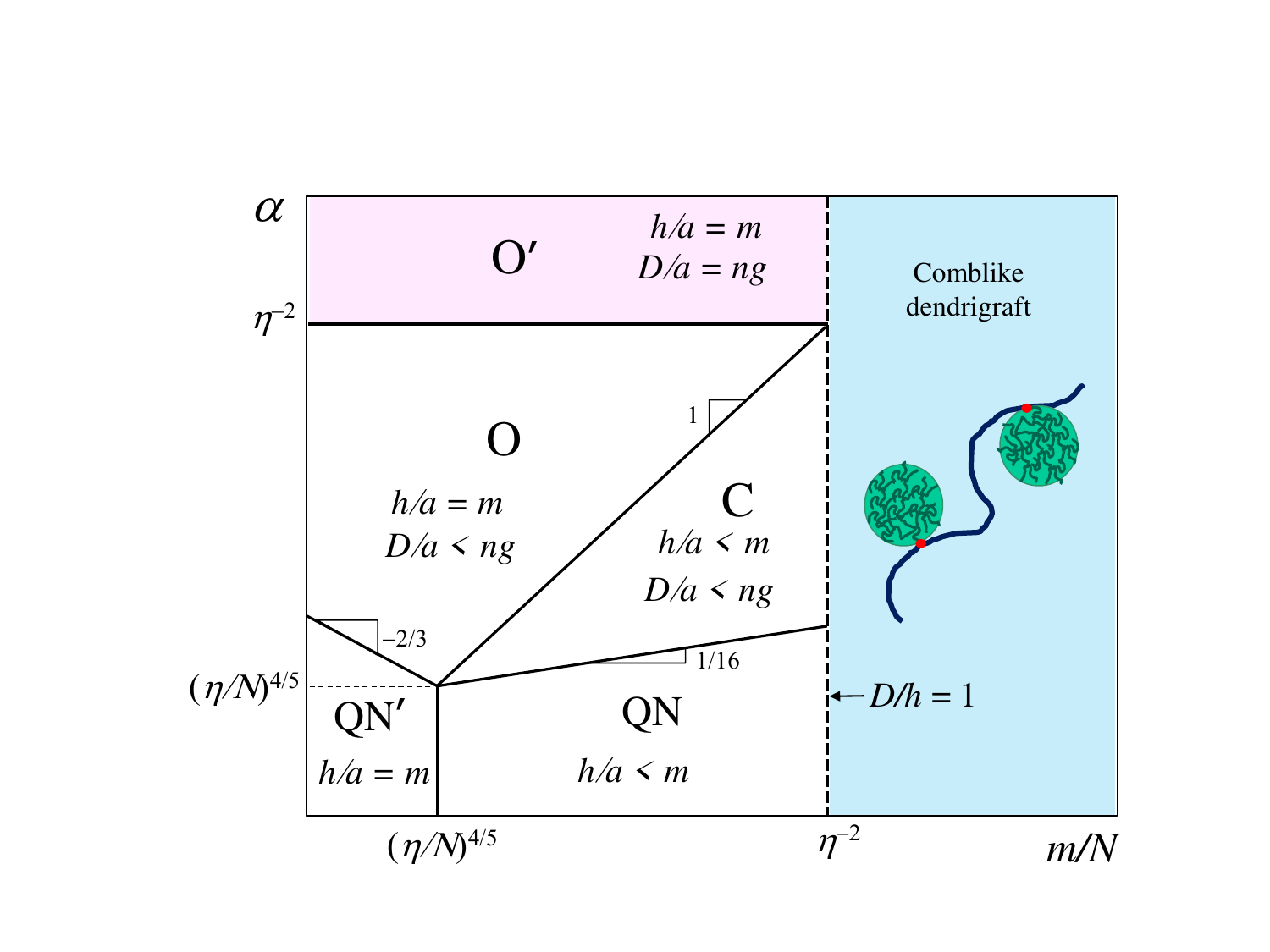}}
\caption{Diagram of states of a dendrigraft in salt-free solution in $\alpha, m/N$ log-log coordinates. 
The Bjerrum length is set $l_B=a$.
}
\end{figure}

\section{Weak polyelectrolyte dendrigraft ("annealing" charges)}

In many practically relevant cases cationic dendrons constituting
dendrigrafts comprise ternary amine groups. They become positively charged
via protonation, and the degree of protonation depends on local pH which in
turn depends not only on the pH in the buffer, but also on salt
concentration and conformation of the dendrigraft.

Let $\beta$ be the fraction of ionizable (basic) monomer units in the
dendrigraft. If fraction $\alpha^{\prime}$ of them is protonated, then the
fraction of elementary (positively) charged monomer units in the dendrigraft
is $\alpha=\alpha^{\prime}\beta$.

If we assume that within intramolecular volume the charge of dendrons is
neutralized by excess number of counterions (osmotic regime), then by
combining mass action law with the local electroneutrality condition (Donnan
equilibrium) the following equation for the degree of ionization of the
dendrigrafts can be derived \cite{APS} 
\begin{equation}
\frac{\alpha ^{\prime }}{1-\alpha ^{\prime }}\cdot \frac{1-\alpha _{b}}{%
\alpha _{b}}=\biggl(1+(\frac{\alpha ^{\prime }\beta c_{p}}{\phi_{s}})^{2}\biggr)%
^{1/2}-\frac{\alpha ^{\prime }\beta c_{p}}{\phi_{s}}  \label{alpha_prime}
\end{equation}%
where 
\begin{equation*}
c_{p}\cong \frac{N}{D^{2}h}
\end{equation*}%
is concentration of monomer units in intramolecular volume of the
dendrigraft, and 
\begin{equation*}
\alpha _{b}=(1-10^{pH-pK_{a}})^{-1}
\end{equation*}%
is the degree of protonation of an isolated ionizable monomer unit in the
buffer with given pH, with $pK_{a}=14-pK_{b}$.

At low salt concentration, $\alpha^{\prime}\beta c_p/\phi_s\gg 1$, the
expansion of the r.h.s in eq \ref{alpha_prime} leads to 
\begin{equation*}
\frac{\alpha'^2}{1-\alpha^{\prime}}\approx \frac{\alpha_b}{1-\alpha_b%
}\cdot \frac{\phi_s}{\beta c_p}
\end{equation*}
which at $\alpha^{\prime}\ll 1$ can be simplified as 
\begin{equation}
\alpha=\alpha^{\prime}\beta \cong \biggl(\frac{\alpha_b}{1-\alpha_b}\cdot 
\frac{\phi_s\beta }{c_p}\biggr)^{1/2}  \label{alpha_prime_1}
\end{equation}

The interaction free energy per dendron can be presented as 
\begin{equation}
F_{int}/k_BT\cong -N(\alpha-\ln(1-\alpha))\approx -2\alpha N\cong - N \biggl(%
\frac{\alpha_b}{1-\alpha_b}\cdot \frac{\phi_s\beta }{c_p}\biggr)^{1/2}
\label{F_ann}
\end{equation}

Minimization of the total free energy with respect to $D$ leads to 
\begin{equation*}
D/a\cong N^{3/2}\eta ^{-2}\biggl(\frac{\alpha _{b}}{1-\alpha _{b}}\phi_{s}\beta
h\biggr)^{1/2}
\end{equation*}%
which can be also formulated as $D/a\cong \alpha N\eta ^{-1}$ with degree of
ionization $\alpha $ given by eq \ref{alpha_prime_1}

By performing subsequent free energy minimization with respect to $h$ or,
equivalently, using eqs \ref{f_h},\ref{f_osm} with $\alpha $ given by eq \ref%
{alpha_prime_1} we obtain the following equation for extension of the main
chain spacers

\begin{equation}
\biggl(\ln(\frac{2x+\sqrt{1+3x^2}}{1-x})\biggr)_{x=h/ma}\cong \frac{\alpha_b%
}{1-\alpha_b}\beta \phi_s N^2\eta^{-2}  \label{h_ann}
\end{equation}
provides dimensions of the dendrigraft in the annealing osmotic regime as

\begin{equation}  \label{h_ann}
h/a\cong m[1-4\exp(- \frac{\alpha_b}{1-\alpha_b}\beta \phi_s N^2\eta^{-2})]
\end{equation}

and

\begin{equation}  \label{D_ann}
D/a\cong N^{3/2}\eta^{-2} \biggl(\frac{\alpha_b}{1-\alpha_b}\phi_s\beta m\biggr)%
^{1/2}
\end{equation}

As follows from eqs \ref{D_ann}, the cross-sectional thickness $D$ , as well
as degree of ionization 
\begin{equation}
\alpha \cong \frac{\alpha _{b}}{1-\alpha _{b}}\beta \phi_{s}Nm\eta ^{-2}
\label{alpha_osm_2}
\end{equation}%
are increasing functions of salt concentration in the osmotic annealing
regime

The cross-sectional thickness reaches its maximal value of $D/a\cong
N(\alpha _{b}\beta )^{1/2}\eta ^{-1}$ at 
\begin{equation}
\phi_{s}^{(max)}\cong (1-\alpha _{b})N^{-1}m^{-1}\eta ^{2}  \label{csmax}
\end{equation}%
when the fraction $\alpha $ of protonated monomer units reaches its maximal
value, $\alpha =\alpha _{max}=\alpha _{b}\beta $.

At higher salt concentration dendrigraft passes into the salt dominated
regime (described in the previous section) with constant value of $\alpha
=\alpha _{max}=\alpha _{b}\beta $. Hence, similar to weak polyelectrolyte
brushes formed by linear polyions, pH-sensitive dendrigrafts are expected to
exhibit non-monotonous dependence of their local structural parameters on
salt concentration, $\phi_{s}$. The position of the maximum $\phi_{s}^{(max)}$ in
this dependence is shifted to higher salt concentration as degree of
branching of the dendrons increases at constant $N$, but is fairly
independent of the number $g$ of dendron generations if $n$ is kept constant.

\section{Numerical self-consistent field modelling }

In order to validate analytical theory predictions concerning response of weak polyelectrolyte dendrigrafts to
variation of salt concentration in the solution we have performed  numerical calculations using 
Scheutjens-Fleer self-consistent field (SF-SCF) method. The detailes of this method can be found, e.g. in ref \cite{Fleer:1993}
and here we show only the results. In our calculations we assumed that the backbone of the dendrigraft is fully extended
(which is a characteristic of osmotic regime) that enabled us to employ one-gradient version of the SF-SCF method.

In Figure 3 we present the salt concentration dependence of the average distance $R_e$ of the end-segments of dendrons from the backbone of the dendrigraft
(the first moment of the distribution)
in the case of the first generation dendrons (arm-tethered stars). 
This structural property can be considered as a measure of crossectional thickness of the dendrigraft $R_e\sim D$.
The functionality of the branching point (the number of free arms) was varied from 3 to 5.
The topological ratio $\eta(q)$ for the first generation dendrons is given by eq \ref{n1}. 
As we can see from Figure 1, the position of maximum in the dendrigraft thickness vs. salt concentration curves remains approximately constant as 
functionality $q$ of the branching point (number of free arms) increases, in accordance with eqs \ref{N}, \ref{n1} and \ref{csmax} which predict
$$
\phi_{s}^{(max)}\sim (q+1) \arctan^2 \frac{1}{\sqrt{q}}
$$
Naturally, the larger is the number $q$ of free arms, the larger is the thickness of the brush in the maximum (cf. eq \ref{D_osm} with $\alpha=\alpha_b$). 

In the insert in Figure 3 d the same dependences are presented in double logarithmic coordinates, the slopes both in low salt and high salt regimes are
in good agreement with predictions of eq \ref{D_ann} and \ref{D_s}, respectively.

\begin{figure}[ht]
\center{\includegraphics[scale=0.50]{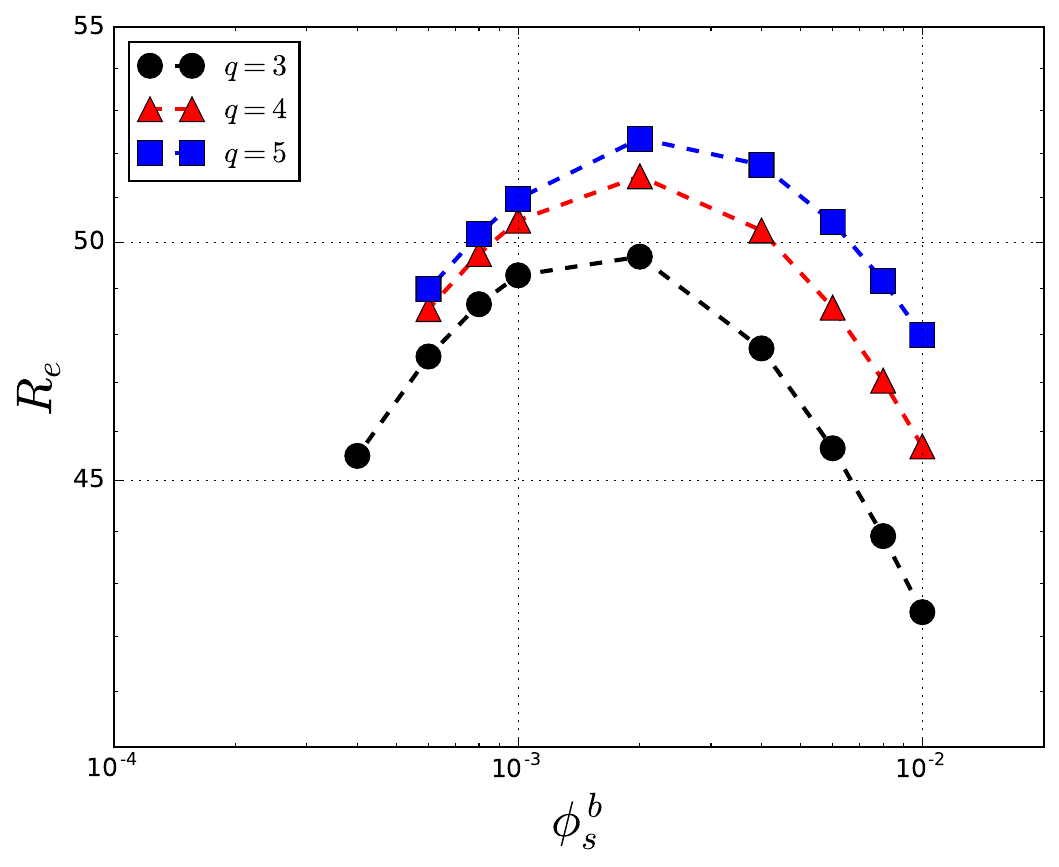}}
\caption{The first moment of the distribution of the dendrons free ends distance from the backbone (average dendrigraft thickness) as a function of salt concentration
for the dendrons of the first generation, $g=1$, and varied number of free arms $q=3,4,5$. Other parameters are: $n=50, h=1$. 
}
\end{figure}

\begin{figure}[ht]
\center{\includegraphics[scale=0.50]{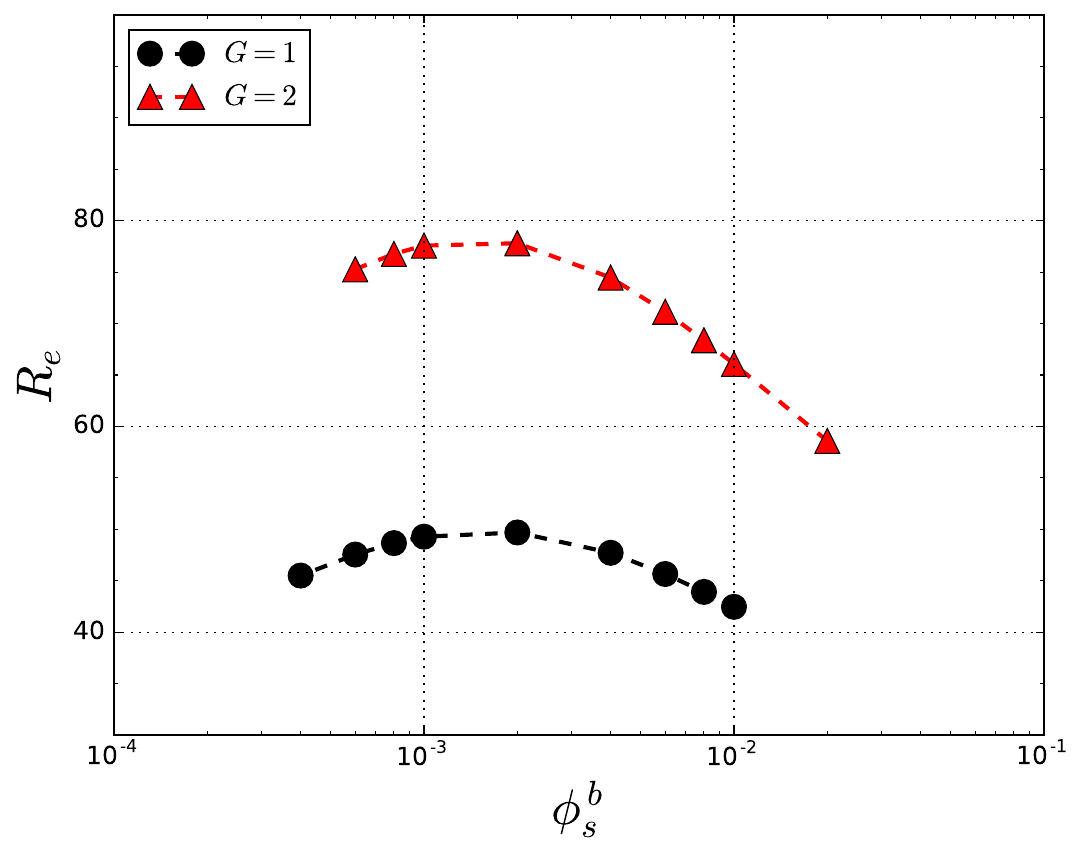}}
\caption{The first moment of the distribution of the dendrons free ends distance from the backbone (average dendrigraft thickness) as a function of salt concentration
for the dendrons of the first $g=1$, and second $g=2$ generation. Other parameters are: $n=50, q=3, h=1$. 
}
\end{figure}

In Figure 4 a similar concentration dependence of the crossectional thickness is shown for dendrigrafts of the 1st and 2nd generations with the same branching functionality $q=3$ and 
the same spacer length $n=50$.
Although the thickness of the dendrigraft of the 2nd generation is noticably larger than that of the 1st generation, the position of the maximum also remains unshifted, as eqs 
\ref{N}, \ref{eta} and \ref{csmax} predict.

\section{Discussion and conclusions}

The considered regimes of dendrigraft behavior are physically similar to the
regimes of  bottlebrushes with linear polyelectrolyte grafts \cite{Borisov_Zhulina_2018}. 
Compared to conventional molecular
brushes, the branched architecture of the grafts (dendrons) manifests itself
in the local properties of dendrigraft via topological parameter $\eta $.
The latter depends on number $g$ of dendron generations and branching
activity (number $q$ of branches emanating from each branching point)

Increasing degree of branching (increasing $\eta $) at constant $N$ leads to
the decrease in dendrigraft thickness $D$, while distance $h$ either
increases or remains constant and equal to limiting extension of spacers of
the backbone, $h\approx am$. 

To illustrate locations of different dendrigraft regimes we present in 
\textbf{Figure 2} the diagram of states for a quenched polyelectrolyte
dendrigraft in $\alpha $, $m/N$ (log-log) coordinates in the salt-free
solution with $l_{B}/a=1$. The diagram contains barely charged (C), osmotic
(O,O') and quasi-neutral (QN,QN') regimes of dendrigraft
with overlapping grafts - dendrons (see \textbf{Figure 1}), separated by
vertical dashed line from "comblike" states with nonoverlapping dendrons
(see schematic in Figure 2). In osmotic sub-regime (marked as O')
tethered dendrons have strongly stretched main path with thickness $D$
approaching its counter length, $D\approx ang\approx aN/\eta ^{2}$, and
spacer with intergraft distance $h\approx ma$. In regime O and 
sub-regime QN' spacers remains strongly stretched with $h\approx ma$, while in
regimes QN and C the stretching of spacers is relaxed, $h<am$. The
boundaries are found from the crossover of dendrigraft parameters, $D$ and $%
h $, in the neighboring regimes. For example, C-O bounadry, $\alpha \simeq $ 
$m/N$, corresponding to onset of counterion condensation in the dendrigraft
volume, ensures crossover of thickness $D$ specified by eqs \ref{D_p_1} and %
\ref{D_osm}. At this boundary the intergraft distance $h\approx am$ (regime
O) starts to relax (regime C, eq \ref{h_p}). All other boundaries (with
slopes indicated) are obtained from similar considerations. The ratio

\begin{equation*}
D/h\simeq \left\{ 
\begin{array}{cc}
\left( \frac{N}{m}\right) \frac{1}{\eta ^{2}} & \text{ regime }O^{\prime }
\\ 
\frac{\alpha ^{1/2}N}{m\eta } & \text{regime }O \\ 
\left( \frac{N}{m}\right) ^{1/2}\frac{1}{\eta } & \text{ \ \ \ \ \ \ regimes 
}C,QN \\ 
\frac{N^{2/3}}{m^{4/3}\eta ^{1/3}} & \text{ \ \ regime }QN^{\prime }%
\end{array}%
\right.
\end{equation*}%
decreases upon a decrease in $m,$ and approaches unity ($D/h\simeq 1$) at $%
m\simeq N/\eta ^{2\text{ }}$(vertical dashed line in figure 2), that is,
when spacer degree of polymerization, $m$, becomes equal (in scaling terms)
to the degree of polymerization $ng$ of the main path of dendron. To the
right of the dashed vertical line ($m>ng$), laterally homogenous brush of
dendrons encompassing the backbone decomposes in nonoverlapping tethered
dendrons, and dendrigraft adopts "comblike" conformation.

If the molecular mass of the dendrons increases exponentially with an
increase in number $g$ of generations according to eq \ref{N}, then by using
eq \ref{eta}, we find that in the salt dominated and quasi-neutral regimes
both $D$ and $h$ increase following the same sub-exponential dependences as
functions of $g$. In barely charged (C) and osmotic (O) regimes $D$ grows as 
$\sim q^{2g/3}$ and $\sim q^{g/2}$, respectively, while $h$ remains
approximately constant. Thus average intramolecular concentration $N/D^{2}h$
either remains constant (in the charged (C) and osmotic (O) regimes) or
weakly decreases (in the salt dominated (S) and barely charged (C) regimes)
as a function of $g$. In annealing osmotic regime both $D$ and $h$ remain
unaffected by an increase in $g$ due to suppressed ionization which leads to
the strong increase in intramolecular concentration of monomer units. These
trends take place as long as dendrons exhibit linear (Gaussian) elasticity,
eq \ref{Fconf_D}, and derived power-law expressions for the dendrigraft thickness $D$ apply.

Local cylindrical symmetry of the dendrigraft is violated near to the ends
of the backbone. At $D\gg h$, which is usually the case, the end-caps of the
dendrigraft can be considered as "stars" formed by $p\cong D/h$ dendrons.
The extension of the dendrons in these hemispherical regions is
approximately equal to their extension in cylindrically-symmetric central
regions of the dendrigraft. The power law dependences for the extension of
dendrons in the end-cap regions can be calculated using similar arguments as
presented above with replacement of average concentration of monomer units
in the interaction term by $c_{p}\cong N/D^{3}$.

\section*{Dedication}
It is our great pleasure to dedicate our work to respectable colleague and very good friend, Matthias Ballauff. 
This theoretical study covers three topics that were in the focus of his own scientific activities during the recent years: dendrimers, polyelectrolytes, and polymer brushes. 
We hope that Matthias will appreciate such collage and simple formulation of physical ideas (e.g., counterion condensation in polyelectrolyte brushes) 
that he persistently pursues and advertises in his experimental research. 
For many years we got inspiration from his experimental works. We always enjoyed and tried to share his curiosity, unlimited enthusiasm and passion in doing science. 
And, of course, we highly appreciate his persistent desire to collaborate with theoreticians, 
even his attempts to cut and replace as many equations as possible by simple physical explanations understandable to everyone.

Ekaterina Zhulina is a Leading Research Fellow at the Institute of Macromolecular Compounds of the Russian Academy of Sciences and 
Professor at Saint-Petersburg State University of Informational Technologies, Mechanics and Optics. 
Current scientific interests include the self-assembly of branched macromolecules in solutions, melts, and at interfaces, inter-polyelectrolyte complexes, 
and associations of polymer-decorated nanoparticles. 

Oleg Shavykin is a Junior Research Fellow at Saint-Petersburg State University of Informational Technologies, Mechanics and Optics. 
His area of expertise covers a wide spectrum of modelling approaches in polymer physics, including Brownian dynamics and self-consistent field
numerical methods. 

Oleg Borisov is a Research Director at CNRS and is affiliated to the Institut des Sciences Analytiques et de Physico-Chimie pour l'Environnement et les Mat\'eriaux, Pau, France. 
His research interests concern polymer and polyelectrolyte brushes, solution properties of branched polyelectrolytes, conformational and mechanical properties of molecular brushes, 
ionic and amphiphilic dendrimers. As Friedrich Wilhelm Bessel Research Award winner he was hosted by M. Ballauff and A.H.E. M\"uller in the University of Bayreuth. 
Co-author of 12 joint publications with Matthias Ballauff on polymer and polyelectrolyte brushes and protein-polyelectrolyte interactions.

\section*{Acknowledgments}

This work was financially supported by Government of Russian Federation
(Grant 08-08) and by the European Union's Horizon 2020 research and
innovation program under the Marie Sklodowska-Curie (grant agreement No
823883).

\section*{Conflict of Interests}
The authors declare that they have no conflict of interest.

\end{document}